\documentclass{IEEEtran}
\usepackage{subfigure}
\usepackage{moreverb}
\usepackage{epsfig}
\usepackage{amsmath,amssymb,amsthm,mathrsfs,amsfonts,dsfont}
\usepackage{adjustbox,lipsum}
\usepackage{algorithm,algorithmic}
\usepackage{amsfonts}
\usepackage{epsfig}
\usepackage{amssymb}
\usepackage{amsmath}
\usepackage{amsthm}
\usepackage{subfigure}
\usepackage{multirow}
\usepackage{rotating}
\usepackage{graphicx}
\usepackage{tabularx}
\usepackage{array}
\usepackage{anyfontsize}
\usepackage{color,soul}
\usepackage{graphicx,dblfloatfix}
\usepackage{epstopdf}
\usepackage{blindtext}
\usepackage{amsthm,amssymb,amsmath,bm}
\usepackage{subfigure}
\usepackage{amsfonts}
\usepackage{epsfig}
\usepackage{amssymb}
\usepackage{amsmath}
\usepackage{cite}
\usepackage{graphicx}
\usepackage{fancyhdr}
\usepackage{subfigure}
\usepackage[subfigure]{tocloft}
\usepackage[font={small}]{caption}
\usepackage{subfigure}
\usepackage{tabularx}
\usepackage{cite}
\usepackage{booktabs,multirow}
\usepackage{tabu}
\usepackage{longtable}
\usepackage[table]{xcolor}

\definecolor{tableHeader}{RGB}{211, 47, 47}
\definecolor{tableLineOne}{RGB}{245, 245, 245}
\definecolor{tableLineTwo}{RGB}{224, 224, 224}

\begin{document}
	\title{{Toward Efficient Transfer Learning in 6G }}

	\author{Saeedeh Parsaeefard and Alberto Leon-Garcia\\Electrical and  Computer Engineering Department, University of Toronto\\ saeideh.fard and alberto.leongarcia @utoronto.ca

	}
	\maketitle
	\begin{abstract}
		6G networks will greatly expand the support for data-oriented, autonomous applications for over the top (OTT) and networking use cases. The success of these use cases will depend on the availability of big data sets which is not practical in many real scenarios due to the highly dynamic behavior of systems and the cost of data collection procedures. Transfer learning (TL) is a promising approach to deal with these challenges through the  sharing of knowledge among diverse learning algorithms. with TL,  the learning rate and learning accuracy can be considerably improved. However, there are implementation challenges to efficiently deploy and utilize TL in 6G. In this paper, we initiate this discussion by providing some performance metrics to measure the TL success. Then, we show how infrastructure, application, management, and training planes of 6G can be adapted to handle TL. We provide examples of TL in 6G and highlight the spatio-temporal features of data in 6G that can lead to efficient TL. By simulation results, we demonstrate how transferring the quantized neural network weights between two use cases can make a trade-off between overheads and performance and attain more efficient TL in 6G. We also  provide a list of future research directions in TL for 6G. 
	\end{abstract}
	\vspace{-.55cm}
	\IEEEpeerreviewmaketitle
	\section{Introduction}
	
	6G will intensify the handling of AI and ML for use cases from various corners of industry and life. 6G will be a network of "integrated intelligence" \cite{HexaX,AIaaSpaper} in which AI/ML algorithms are applied for autonomous networking and where the network will be redesigned to support the AI/ML based applications \cite{ITU2030,ITUTreport,3GPP2,ENI}. 
	
	The success of AI/ML algorithms requires large volume, updated data sets with similar statistical conditions in the training and testing phases \cite{8761327}. However, this is not the case for many real scenarios for two main reasons. First, collecting data and providing labeling are not always accessible (e.g., cyber attacks) and may involve high cost. Second, dynamic environments in real scenarios call for updating data sets regularly (e.g., fast and dynamic nature of the wireless channel and uncertainty in the behavior/mobility patterns of end users in intelligent transportation systems (ITS)). Re-collecting  data in these scenarios and rebuilding  models from scratch impose huge burdens for these use cases.

	Transfer learning (TL) or knowledge transfer (KT) is a promising approach to handle both of the above-mentioned challenges \cite{5288526,Zhuang2019ACS,8761327,DBLP:journals/corr/abs-2102-07572}. In this context, it is possible to transfer lesson-learned parameters (knowledge) among learning algorithms that includes data sets, models, hyper parameters, and optimization approaches. In particular, each learning algorithm can be divided into two parts: 1) a domain containing feature data and its statistics; 2) a task involving models and labels. Typically in TL there are at least one source and one target AI/ML algorithms in TL where the knowledge of domain and/or task of the source is transferred to a target.

	Experience with TL shows that sharing domains and tasks can improve the speed of convergence and learning accuracy. this case will be refereed to "a positive TL". However, when there is not enough similarity and correlation among domains and tasks of the source and the target,  TL might have negative effect on the performance of the target. 
	Applications of TL in 6G and its positive potential in future networking has recently drawn a lot of attention, e.g.,  \cite{9388790, 6Gurrllcsurvey}. In this paper, our goal is to study how we can attain  efficient TL in 6G.

	We begin by looking at the TL categories. Then, we explain how 6G architecture planes can be modified to provide the efficient TL. We discuss the deployment of TL pipe-lines in the infrastructure, novel procedures in management plane to handle TL among sources and targets, new repositories and knowledge extraction modules in a training plane, and modifications in application plane. Based on the network layers, we provide examples about data sets and models that may be saved in repositories of the training plane in 6G. 
	
	Fortunately, in wireless networks, there exist spatio-temporal and hierarchical correlations among AI/ML algorithms running in different nodes that are conducive to positive and efficient TL. Despite its great potential, TL in 6G introduces various implementation challenges. For example, use cases  can impose diverse granularity requirements of TL in 6G involving real time, online/offline, and on-demand TL. And there are concerns about robustness, reliability, security, and privacy of pipe-lines for TL. We will discuss these implementation challenges in 6G and highlight the overheads of TL in 6G. In an evaluation section, we show how quantization approaches can be used with TL to reduce the extra overhead inside of 6G.  
	
	The organization of this paper is as follows. In Section II, we briefly review TL. In Section III, we show how TL changes the 6G architecture. Section IV contains future research directions. In Section V, a deployment scenario for TL in 6G will be discussed; followed by conclusions in Section VI.

	\section{Transfer Learning Preliminaries}
\begin{figure}
		\begin{center}
			\includegraphics[width=3.2 in]{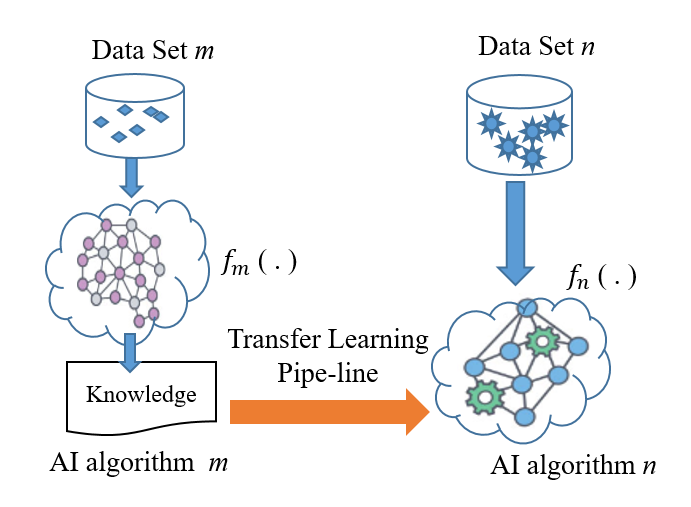}
			\caption{Transfer learning and its pipe-line inside of 6G }
			\label{TL3}
		\end{center}
	\end{figure}

	Consider an AI/ML algorithm $m$ that has  \cite{5288526}: 
	\begin{itemize}
		\item \textit{Domain}: consisting of a  feature space $\mathcal{X}_m$ for algorithm $m$ and a marginal probability distribution $P_m(\textbf{X}_m)$, where $\textbf{X}_m = \{x^1_m,\cdots,x^n_m\} \in \mathcal{X}_m$, which constitute the domain $\mathcal{D}_m=\{\mathcal{X}_m,P_m(\textbf{X}_m)\}$ 
		\item \textit{Task}: involving a label space $\mathcal{Y}_m$ and a predictive function $f_m(\dot)$ which shows the relation between $\mathcal{X}_m$ and  $\mathcal{Y}_m$, and the task is denoted by $\mathcal{T}_m=\{\mathcal{Y}_m, f_m(.)\}$. 
	\end{itemize}
	In supervised learning mode, the data set $m$ consists of $\mathfrak{D}_m=\{(x_m^1, y_m^1), \cdots (x_m^n, y_m^n)\}$ and $f_m(.)$ is found by minimizing the distance between the feature and the label spaces. In unsupervised learning mode, the task changes to $\mathcal{T}_m=\{ f_m()\}$ since the labels are not available and the best predictive function is matched to the feature space. Traditionally, the domain $\mathcal{D}_m$ and the task $\mathcal{T}_m$ are fully disjoint from any other $\mathcal{D}_n$ and $\mathcal{T}_n$ for $m \neq n$. However, in TL, parameters among domains and tasks $m$ and $n$ for all $m$ and $n$ can be shared, see Fig. \ref{TL3}. Any shared part of the domains or the tasks can be referred as "knowledge", e.g., all or the portion of the feature and the label spaces (data sets), or $f(.)$ and its hyper-parameters (models). TL aims to improve the performance of $\mathcal{T}_n$ with $\mathcal{D}_m$ and $\mathcal{T}_m$. We consider the following two key performance measures for target $n$:
	\begin{itemize}
		\item \textit{Accuracy, consistency and generalization} of learning in the target domain:  $\eta^{TL}_n=\frac{P^{\text{TL}}_n}{P^{\text{Traditional}}_n}$ where $P^{\text{TL}}_n$ and $P^{\text{Traditional}}_n$ are the performance measures of the AI algorithm $n$ with and without TL, respectively.  
		\item Training time in the target domain: $\tau^{TL}_n=\frac{T^{\text{Traditional}}_n}{T^{\text{TL}}_n}$ where $T^{\text{TL}}_n$ and $T^{\text{Traditional}}_n$ are the training times of the AI algorithm $n$ with and without TL, respectively. 
	\end{itemize}
	Now we can defined positive TL as a case that $\eta^{TL}_n>1$ and $\tau^{TL}_n>1$. The above definitions are for one source and one target and it can be extended to any set of sources and targets. 
	
	From 6G's preservatives, to have successful TL, required bandwidth, delay, the level of security and privacy of passed knowledge between sources and targets should be determined. Also, the granularity of TL algorithm and the iteration number to pass knowledge between source and target should be determined. For instance, we can have following classes of interaction among source and targets:  
	\begin{itemize}
		\item \textit{Real-time TL} where the knowledge is passed between source and target as soon as a new update is available; 
		\item \textit{Non real-time or semi-real time TL }where the knowledge  between the sources and the targets are passed according to a schedule; 
		\item\textit{ On-demand TL} where the knowledge is passed from the source to the target when a request from the target is initiated. 
	\end{itemize}
	These types of interactions together with the end-to-end (E2E) delay determine the required bandwidth between each source-target pair. These parameters can be used to deploy the pipe-line for TL, demonstrated in Fig. \ref{TL3}. An important  point is that the TL pipe-line causes additional overhead and cost for passing knowledge. Below we identify different types of overheads and costs are incurred in TL. One approach to obtaining a single measure for overhead and cost is to use a weighted sum of the components: \begin{eqnarray}
		\lefteqn{\Theta^{\text{TL}}_{mn}=\alpha_1 f_1(W_{mn})+ \alpha_2 f_2(D_{mn})}\\&& \,\,\,\,\,\,\,\,\,\,\,\, +  \alpha_3 g_{mn}(\text{ interaction}) +\alpha_4 h_{mn}(\text{security}), \nonumber
	\end{eqnarray} where $\alpha_1, \alpha_2, \alpha_3, \alpha_4$ are regularization weights and functions are defined as:
	\begin{itemize}
		\item $f_1(W_{mn})$ and $f_2(D_{mn})$ are increasing and decreasing functions of required E2E bandwidth $W_{mn}$ and delay $D_{mn}$, respectively;
		\item $g_{mn}(\text{ interaction})$ shows the cost per interaction class, defined as
	$$g_{mn}(\text{ interaction}) = 
		\left\{\begin{array}{lr}
			M_1, & \text{if} \,\, \text{Real-time},\\
			M_2, & \text{if } \,\,\text{Non-Real time},\\
		M_3  &  \,\,\text{Otherwise,}
		\end{array}\right.$$
		where $M_1$, $M_2$, $M_3$ are positive values demonstrating the costs related to each class and they can be set as $M_1\geq M_2 \geq M_3$ by wireless network infrastructure providers. The good policy also can be designed to reduce the cost of overhead, for instance by sending the TL knowledge among the source and the target in non-rush hour time windows. 
		\item $h_{mn}(\text{security})$ is the security related cost function and it is an increasing function with the level of requested by source $m$ and target $n$. 
	\end{itemize} 
	Clearly, decreasing the overhead while attaining positive performance of the learning AI are major indicators of effectiveness of TL in 6G. 
	
	There are several categories of TL based on \textit{what} can be transferred among sources and targets and the similarities and differences between their domains and tasks. For example, based on the availability of labels, there are three categories of TL:
	\begin{itemize}
		\item Transductive TL where the label data is available only from the source not the target and $\mathcal{D}_m\neq \mathcal{D}_n$ while $\mathcal{T}_m = \mathcal{T}_n$
		\item Inductive TL where label data are available for the source and targets but the tasks are different, i.e., $\mathcal{T}_m \neq \mathcal{T}_n$
		\item Unsupervised TL where the label data is not available in neither source or target
	\end{itemize}  
	Based on domains, we have 
	\begin{itemize}
		\item Heterogeneous TL where $\mathcal{D}_m \neq\mathcal{D}_n$ 
		\item Homogeneous TL where $\mathcal{D}_m=\mathcal{D}_n$ 
	\end{itemize}
	and based on the solution approach, we have 
	\begin{itemize}
		\item Instance-based TL where samples of data sets are shared between the source and target
		\item Feature-based TL where original features of the source are matched to target ones, or where for both source and target features, latent features or common features are discovered. 
		\item Parameter-based TL which involves transferring the parameters of models 
		\item Relational-based TL which transfer the learned rules from source to the target   
	\end{itemize}
	In the following, we will discuss how TL should be processed in 6G and give examples of the above categories of TL in 6G.

	\section{Aspects of Efficient TL in 6G}
	In this section, we study effects of TL in 6G and show how 6G should be adapted to serve AI/ML use cases. Many AI/ML use cases in 6G involve learning/control loops containing \textit{monitoring, analysis, policy, execution plus knowledge} steps, denoted as (MAPE-K) loops. A MAPE-K loop provides a unified way to represent these AI/ML based applications \cite{ENI,AIaaSpaper}. TL can be deployed between ML algorithms that are normally present in the analysis step of MAPE-K loops associated diverse services in OTT, and/or networking  applications (NAL). In this paper, for simplicity, we assume there is one MAPE-K loop per each AI/ML algorithm which can be a target or a source of TL. Let's start our discussions with some examples to show how TL can be used in 6G.

	\subsection{Examples of TL in 6G}
\begin{figure}
		\centering
		\includegraphics[width=2.9 in]{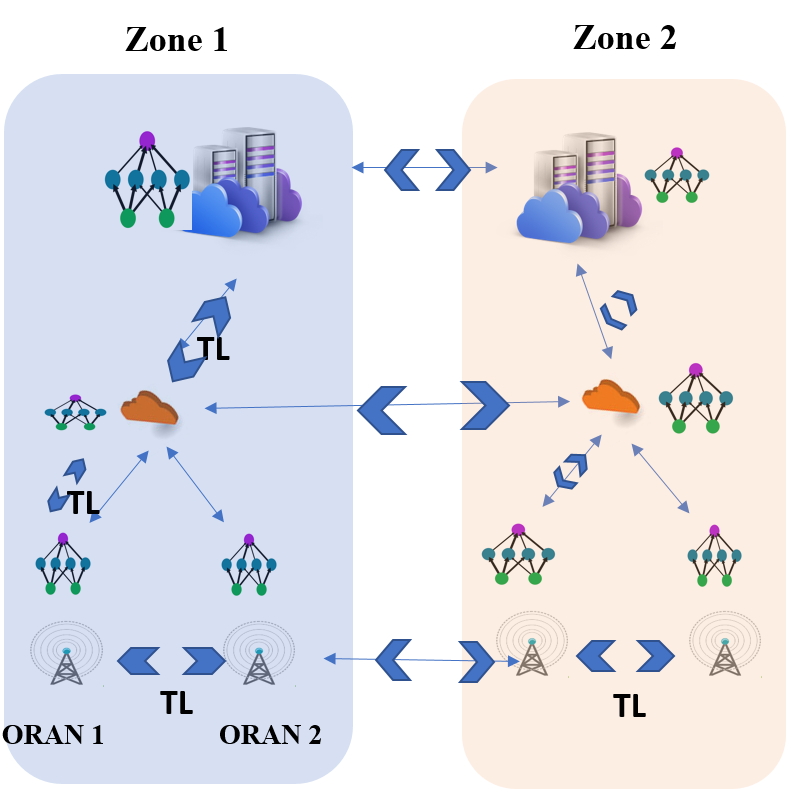}
		\caption{One scheme of the TL examples in zone 1 and zone 2 of 6G}
		\label{fig:TL4}
	\end{figure}

	In 6G, many use cases in networking share the spatio-temporal features or similar information in performing various tasks. Let's focus on resource allocation (RA), energy efficiency (EE), and admission control (AC) of radio access networks (RANs) for one specific area in 6G. There are several access nodes, e.g., ORANs, in a region that provide coverage (see Fig. \ref{fig:TL4}).
	
	All of these tasks utilize the mobility pattern and traffic pattern of end-users. Therefore, the domains are similar but the tasks are different. Any ORAN can share its own domain among its own different tasks. Also, ORAN 1 can share its own task and domain for the resource allocation (RA) with the resource allocation (RA) in ORAN 2; which is an example of to transduction TL where the source is RA in ORAN 1 and the target is RA in ORAN 2. If the resource allocation (RA) in RAN 2 also shares its domain or tasks with the resource allocation (RA) in RAN 1, we have inductive TL. If both RANs do not have labeled data, there is unsupervised learning among them. The resource allocation (RA) in ORAN 1 and ORAN 2 can share their knowledge in real-time, non-real time and on-demand cases. For the last two cases, depending on the amount of available bandwidth, TL among ORAN 1 and ORAN 2 can be scheduled  to reduce $\Theta^{\text{TL}}_{\text{ORAN1}, \, \text{ORAN2}}$. The same is true for the energy efficiency (EE) in both ORAN 1 and ORAN 2. However, there is a chance that the energy efficiency (EE) decision is in contrast with an RA decision. For instance, the resource allocation (RA) requires more resource to guarantee the QoS while the energy efficiency (EE) asks for less power consumption and releasing some resources. These examples demonstrate the importance of interaction and TL management in 6G. 
	
	These AI/ML algorithms can share the domains and tasks with another part of the RAN in different cities or different locations. In this case, we can have parameter or instance or feature based TL. The resource allocation, admission control, and the energy efficiency in ORANs of one area can share their models and data sets with the core elements to derive more large scale mobility patterns and network user behaviors. In this case, one probable scenario is to share the domain of RANs with the core to increase the size of data sets of core elements without additional cost. The knowledge can be fully or partially shared among AI/ML algorithms in 6G. 
	
	The mobility pattern of 6G nodes, which arises from NAL applications, can be transferred to ITS, which is an OTT application,  or vice versa. Various types of TL can be applied here. The similarity between these two use cases is obvious. However, 6G can encounter more complex scenarios. For example, for cyber-security, 6G may require a behaviour and traffic model of legitimate users of OTT use cases. In this case, the domains and the tasks of both cases are not correlated. Therefore, we have heterogeneous TL here, and any solution based categories of TL can be applied.

	\subsection{Hierarchy and Spatio-temporal Features of 6G and TL}
\begin{figure}
		\centering
		\includegraphics[width=3.62 in]{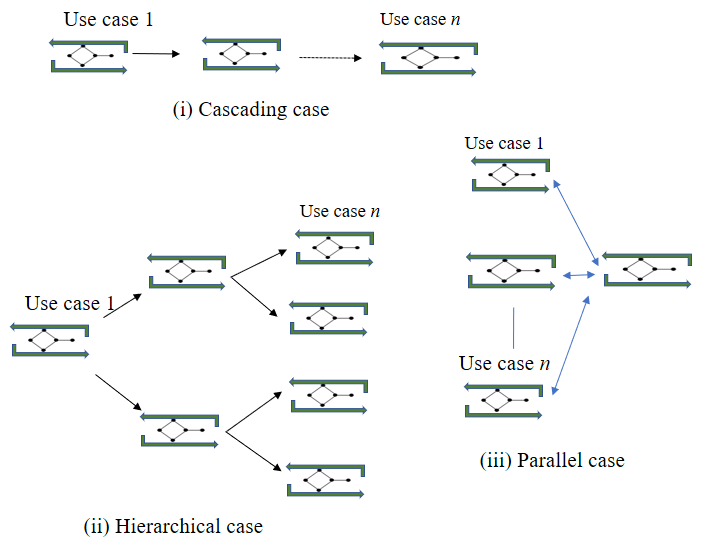}
		\caption{The hierarchy and interaction among AI/ML use cases in 6G for three models of interaction (i) cascade model (ii) Hierarchical model, (iii) parallel model. }
		\label{fig:TL2}
	\end{figure}

	Wireless networks have hierarchical structures. Also, our examples in the previous section show that there is an inherent spatio-temporal feature among different location zones, RANs, and core entities. These features  help create positive TL in 6G for clustering and classifications AI/ML algorithms. Therefore, hierarchy and Spatio-temporal features of traffic of users in 6G can be used as the criteria for the clustering AI/ML algorithms for TL. In Figs. \ref{fig:TL2} (a) and (b), we plot some models of interaction of  AI/ML algorithms for TL in 6G which can basically derived from both hierarchy or spatio-temporal feature of 6G. In Fig. \ref{fig:TL2}, we present three models: 
	\begin{itemize}
		\item Cascade AI/ML algorithms in (i) which can show the neighboring ORANs transferring their models and data sets for mobility or traffic patterns of users
		\item Hierarchical AI/ML algorithms in (ii) which can represent the interaction between access and core elements in TL 
		\item Parallel AI/ML algorithms (iii) belong to the case that the similar nodes (base stations, switches, storage devices) collaborate to learn the similar models based on their local data sets through the concept of federated learning (FL). Also, they can share their public (not very sensitive data sets in terms of privacy) with their peers. 
	\end{itemize}
	This hierarchical structure inside of 6G can help to find the correlation, similarities, and relation among sources and targets. 
	
	\subsection{6G Architecture Modifications}
	
\begin{figure}
		\begin{center}
			\includegraphics[width=3.1 in]{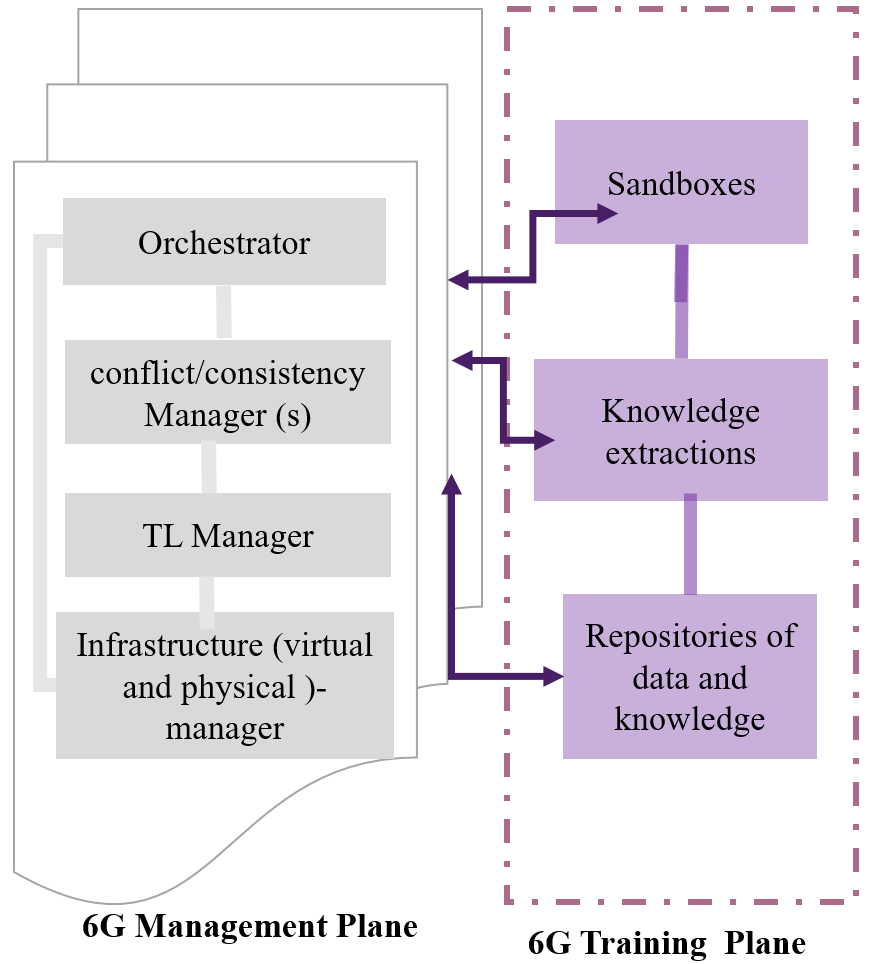}
			\caption{Training and management aspects of 6G to provide TL.  }
			\label{6Gtrainingmanagement}
		\end{center}
	\end{figure}

	We see the 6G architecture as including four planes to serve these use cases and to handle their needs associated with TL \cite{9200631,AIaaSpaper}. We define responsibilities of each plane as follows:
	\begin{itemize}
		\item \textbf{Application Plane} which represents the MAPE-K loops as chains of functions/services that implement the data collection processing and action pipe-lines that implement the loops, along with their required QoS and QoE. The ML algorithm(s) in each MAPE-K loop can be the source or target for algorithms in other MAPE-K loops. Application plane determines following items for TL:
		\begin{itemize}
			\item A list of trustful targets; 
			\item A list of trustful sources;
			\item Parts of models and tasks that are allowed to be shared; 
			\item Required levels of security and privacy;
			\item TL angularity with any specific source and target; 
			\item Privacy and security level;
			\item Required E2E delay for TL with any specific source and target;
			\item Required bandwidth for any allowable TL.
		\end{itemize}
		The above parameters help to determine the service level agreement (SLA) for any pair of the source and target in TL. These SLAs will be used in 6G to establish appropriate pipe-lines per each MAPE-K loop. 
		\item \textbf{Infrastructure Plane} which is an E2E cloud-native, multi-tier, programmable structure with virtualized communication and computation resources. This plane inherits the software defined and virtualized structures of 5G and it is responsible to provide TL pipe-lines from sources and targets for both OTT and NAL applications. The request for setting up a TL pipe-line is sent by a management plane to the infrastructure plane. Other pipe-line parameters are derived from SLAs in application plane. 
		\item \textbf{Management Plane} which is in charge of monitoring all running services and Required admission for new requests of services in fully autonomous manner. In 6G, this plane should include procedures to handle the conflicts that may arise among MAPE-K loops, pertaining to the consistency and coherency of loops \cite{AIaaSpaper}, and TL management. These new procedures along with orchestrator and infrastructure management are shown in boxes in Fig. \ref{6Gtrainingmanagement}. TL manager has following tasks:  
		\begin{itemize}
			\item Determining who can initiate TL among MAPE-K loops and services in 6G. For instance, among networking applications, TL request can be issued by network elements (decentralized manner) or can be handled by the orchestrator (centralized manner). For the OTT applications, the request can be originate with the OTT applications; 
			\item Determining authorized TLs among MAPE-K loops; 
			\item Initiating and handling TL for one or multiple service/infrastructure providers and OTT slices for different organizations. 
		\end{itemize}
		The request for setting up TL pipe-line among MAPE-K loops can be initiated from use cases. For example in an OTT, an Intelligent Transportation System (ITS) may send a request for TL to get data about the mobility pattern of users. In a NAL setting, a request for TL can be started by network management plane. In both cases, the orchestartor will handle the setup procedure for providing the TL pipe-line in the infrastructure by monitoring security, reliability, and privacy. In general, all the \textit{coordination} among sources and targets are handled by this plane. This task is more complex when dealing with  multi-source multi-task TL and when the sources and targets are in different networks or OTT applications. 
		
		\item \textbf{Training Plane} which is  a new plane in 6G to integrate AI and ML into the network  \cite{AIaaSpaper,ITUTreport} and to provide sandboxes that mimic the real environment with the goal of  training, retraining and examining any AI based use cases in offline, near-online and online manner  \cite{ITUTreport}. To support TL, this plane provides an environment to measure the TL positive effects on the targets' learning procedures. It should learn the similarity and correlations among sources and targets to attain positive TL for both NAL and OTT use cases. It also needs to be equipped with the means to extract the knowledge from domains and tasks of each AI based application. We represent these tasks in four boxes in Fig. \ref{6Gtrainingmanagement}, defined as: \\1)
		\textit{Sandboxes} which are responsible to provide simulation/ emulation environments for TL, clustering and finding the beneficial correlation among sources and targets. These reports should be sent to the second box of the management plane. \\2) \textit{Knowledge extraction } which can provide parameters from domains and tasks of sources and targets for TL (knowledge). This knowledge is ready to transfer between source-target pairs depending on TL categories. \\3) \textit{Repositories of data and knowledge} which are storage units to store and retrieve  knowledge for TL among sources and targets. 
	\end{itemize}
	In 6G, the training and management plane have tight interaction to attain efficient TL. All actions in boxes in the training plane are sent to the management plane to execute the interaction and knowledge transfer among the MAPE-K loops. In Fig. \ref{6Gtrainingmanagement}, we show arrows between boxes to highlight the importance of these interfaces. 

	\subsection{Repository of Knowledge of TL in 6G}
	In the repository, the models and data sets of each target-source pair can be stored and used for TL among similar/correlated OTT and networking use cases when permitted. For instance, the following information can be stored in the repository:  
	\begin{itemize}
		\item End-Users which are smart phones, IoT devices (e.g., smart watches), autonomous vehicles, and sensors. These devices can share the following: 
		\begin{itemize}
			\item Models and data sets related to UE/Session related parameters, e.g., mobility pattern, spatio-temporal traffic model, wireless channel and coverage model, QoS, and QoE. 
		\end{itemize}
		\item Radio-access level devices including base stations, radio remote heads (RRHs), cloud radio access networks (CRAN) or open RANs (ORANs) where they provide coverage for specific area. These entities can share the following:  
		\begin{itemize}
			\item Models and data sets per each region for spatio-temporal traffic and mobility models of users in that area, network measurements and logs, RAN-related parameters (e.g., channel quality), load, interference, user density. resource condition parameters (e.g., availability of front-haul), reliability and robustness of services.   
		\end{itemize}
		\item Core-level devices including switches, firewalls, packet gateways, computing devices, storage devices of the users' information, and clouds inside of the networks, which can share the following:  
		\begin{itemize}
			\item Models and data sets for processing load and availability of core functions (both virtual and physical), load and availability of computing resources, large scale spatio-temporal traffic and mobility per zone for each type of services and slices, core network measurements and logs, and procedure level task information.
		\end{itemize}
		\item OTT and service applications including traditional data and voice services, smart services for different sectors of industry, and slices which can provide 
		\begin{itemize}
			\item Models and data sets for slice QoS and QoE, slice isolation, spatio-temporal traffic and mobility per slice, measurements, and logs per slices, 
		\end{itemize}
		\item Management level which includes all the OSS and BSS parameters  and they can share the following parameters inside of the network or with another network provider: 
		\begin{itemize}
			\item Models and data sets for fault, configuration, accounting, performance, and security (FCAPS) and the management reports 
		\end{itemize}
	\end{itemize}
	Repository of knowledge in TL can be handled by training plane where seasonal, daily, and weekly patterns of any knowledge, models and data sets can be provided for different MAPE-K loops. However, the management plane takes care of initiating and allowing the TL procedures between any source and target in private and secure manner.

	\section{Challenges and Future Works }
	TL will play an important role in simplifying and streamlining the use of AI and ML in 6G networks. One of the major challenges of deploying TL is to deal with the vast number of possible approaches and formulations, so choosing the best TL approach is a major implementation challenge in 6G. Also, TL mechanism should be tuned based on the use case which poses the challenge of finding new procedures to handle TL in more general manner.  
	
	There exist overlaps between TL and other concepts such as federated learning, semi-supervised learning, multi-view learning, and most importantly, multi-task learning. Having a unified view of TL and highlighting its overlaps with other context can unlock the potential of TL in 6G. 
	
	Adaptation of TL for reinforcement learning (RL) in dynamic environment is another challenge \cite{zhu2021transfer,JMLR09-taylor}. Here, to have positive TL, transformation of different parts of the information from the source (e.g., state-space, action-space, reward, and transition dynamics) to the target should be studied. Since there is more spatio-temporal dynamic behavior in the system, TL needs more adaptation of the parameters. Most of the recent works are well-tuned approaches for specific applications of TL in RL. Framework-agnostic TL for RL is of desirable in 6G. 
	
	Positive TL requires appropriate clustering, correlation, and similarity detection among sources and targets. This can be efficiently handled for homogeneous TL among sources and tasks. However, in heterogeneous scenarios, adapting the source and target domains and tasks are not trivial. Deriving latent variables in data sets of different AI/ML algorithms can also lead to positive TL and attain the better clustering and similarities among tasks. 
	
	In the most works, TL is used for a pair of one source and one target. However, as mentioned in the examples above, in 6G, there is a set of sources aiming to share their information with a set of targets. Expanding all the existing formulations of one source-target pair TL to multi-source multi-target TL in 6G can be considered as another future research direction.  
	
	Preserving the required security and privacy and preventing any threat from attackers during the knowledge transfer among AI/ML algorithms are also another important issue for efficient TL in 6G. TL in 6G should be immunized against eavesdropping, GAN attacks (for models and data sets), poisoning, and Sybil attacks.

	\section{Deployment Scenario:\\ Quantization to Reduce TL Overhead}
	To overcome TL extra overhead in 6G, in this section, we study how quantization algorithms can compensate this practical challenge. Quantization in this context is a class of techniques for performing computations and storing tensors at less number of bits than floating point case. Consequently, a quantized model executes some or all of the operations on tensors with integers rather than floating point values, leading to a more compact model representation for DNNs. 
	
\begin{figure}
		\centering
		\includegraphics[width=0.45\textwidth]{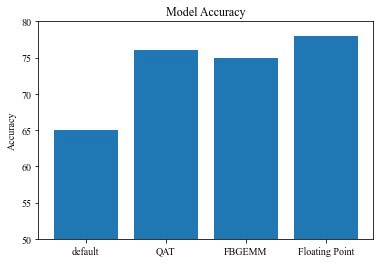}
		\caption{Percentage of accuracy of target versus quantization approaches}
		\label{fig:quantization-acc}
	\end{figure}

\begin{figure}
		\centering
		\includegraphics[width=0.45\textwidth]{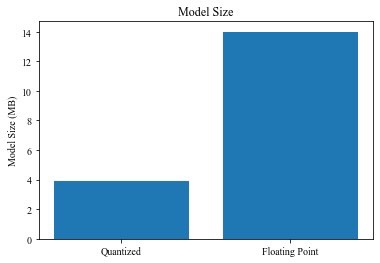}
		\caption{The amount of overhead of quantized versus non-quantized TL in 6G}
		\label{fig:quantization-size}
	\end{figure}

	We implement TL pipe-line between two neural networks in
	which one of them is trained on the CIFAR-10 data set and the neural network model is
	MobileNetV2 \cite{mobilenet} and we apply \cite{quantization}.  Fig. \ref{fig:quantization-acc} shows the performance of quantized algorithms with the case that the exact values of weights are passed
	among neural networks, i.e., \textit{Floating point method} which requires 32 bits of
	data per weights of neural networks. In
	Fig. \ref{fig:quantization-acc}, the accuracy of training phase shows the
	performance of floating point approach versus following quantization algorithms
	for 8 bits per each weight \cite{quantization}: 
	\begin{itemize}
		\item \textit{Default} quantization  approach which applies uniform
		quantization levels between minimum and maximum values of weights
		\item \textit{FBGEMM} where by considering the
		distribution of the weights, quantization levels is adjusted
		which leads to better
		accuracy compared to \textit{Default} quantization  approach 
		\item \textit{Quantization aware training (QAT)} In the above
		approaches, quantization is applied at the end of the model training.
		In this approach, after that, the model is fine-tuned on the quantized weights by
		training a couple of more iterations to improve the accuracy. In our experiments,
		this method leads to 1\% better accuracy compared to \textit{FBGEMM} from Fig. \ref{fig:quantization-acc}.
	\end{itemize}
	Fig. \ref{fig:quantization-acc} highlights that by 75\% decreasing the load of the network (8 bits/32 bits) ($f_{mn}(\text{interaction})$), still the accuracy of training phase is in the acceptable range for \textit{QAT}. To demonstrate it figuratively, in Fig.
	\ref{fig:quantization-size}, the size of the quantized model is compared to that
	of the original model. As shown in this figure, the size of the model is reduced
	to 75\%. This simulation shows that the quantization is an appropriate
	method to make a trade-off between performance and amount of messages in the AI-aaS based networks. 
	
	After passing the quantized weights, we can re-tune the final layer of the model in the target to increase the accuracy. We use a Pytorch tutorial of "Quantized Transfer Learning for Computer Vision" and its colab version  \cite{quantizationforTL}. By re-training the model at the target, the accuracy level will be increased up to the case without quantization. The re-training phase in the target takes $3$ m and $16$ sec versus while full training requires $10$ m and $16$ sec for convergence. On the other hand, we have $\tau^{\text{TL}}=3.142$, meaning that TL can improve the energy efficiency with the factor $\tau^{\text{TL}}$, and consequently, improving sustainability in 6G.
	
	\section{Conclusion }
	In this paper, we address how efficient transfer learning (TL) in 6G can be attained. We introduce metrics to assess efficiency TL from both learning algorithm and networking aspects. We discuss possible modifications in 6G architecture to handle TL and we define new elements in the network management plane. We focus on new activities in training plane and its repositories for TL in 6G. With some examples and looking at spatio-temporal features of wireless networks, we show how classifications and clustering of algorithms in 6G can be attained to reach efficient TL. For positive TL, we require more unified view of TL approaches with security and privacy guarantees discussed in this paper.

	\bibliographystyle{IEEEtran}
	
	\bibliography{IEEEabrv,myref}

\end{document}